\documentclass[prb,preprint]{revtex4}
\usepackage{epsfig}

\begin{document}

\title{How much can donor/acceptor-substitution change 
the responses of long push-pull systems to DC fields?}

\author{Michael Springborg$^a$ and Bernard Kirtman$^b$}

\affiliation{$^a$Physical and Theoretical Chemistry, University of 
Saarland, 66123 Saarbr\"ucken, Germany
\\ $^b$Department of Chemistry and Biochemistry, University of California, 
Santa Barbara, California 93106, U.S.A.}

\date{\today }

\begin{abstract}

Mathematical arguments are presented that give a unique answer to the question in the title.
Subsequently, the mathematical analysis is extended using results of 
detailed model calculations
that, in addition, throw further light on the consequences of the analysis. 
Finally, through a
comparison with various recent studies, many of the latter are given a new interpretation.

\end{abstract}

\maketitle

\section{Introduction}
\label{sec01}

The response to 
external electric and magnetic fields provides a fundamental tool for studying and altering
the properties of materials with numerous attendant applications. In particular higher-order
responses allow for `manipulating light with light.' Thus, there is considerable
interest in identifying molecular systems with large non-linear responses. 

\begin{figure}
\begin{picture}(15,10)
\put(00,0){\psfig{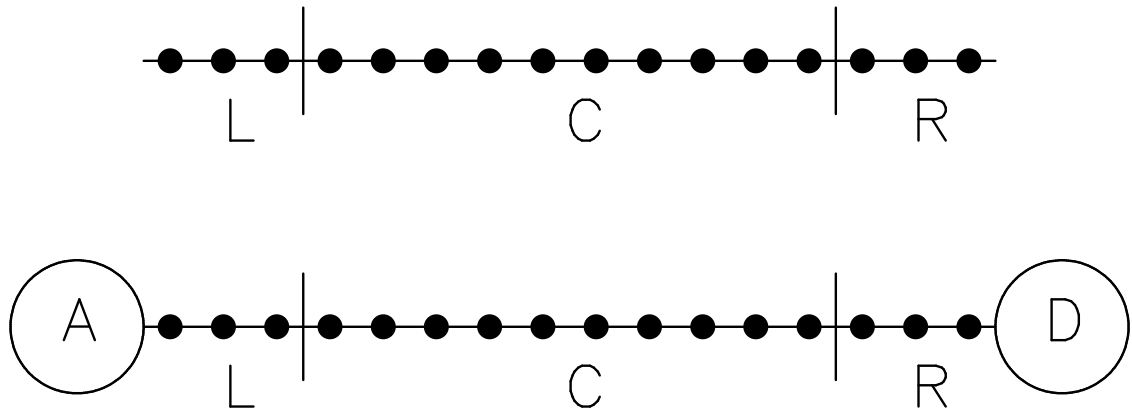}}
\end{picture}
\caption{Schematic representation of a long finite chain separated into
a central region and two terminal regions. The lower chain shows what 
happens when a donor and an acceptor group have been added at the terminations.
Each dot represents a group of atoms.}
\label{fig01}
\end{figure}

One approach in this direction is based on push-pull systems, i.e., chain-like 
molecules with an
electron-donor group at one end and an electron-acceptor group 
at the other (see Fig.\ \ref{fig01}). When the backbone is a $\pi$-conjugated oligomer
the $\pi$ electrons of the backbone may respond easily to perturbations like those of the 
substituents and/or
external fields. Due to the donor and acceptor groups a large electron transfer, 
and, accordingly, a large dipole moment can occur and one may hope for large
responses of the 
dipole moment to external fields. For these $\pi$-conjugated systems, each circle in 
Fig.\ \ref{fig01} could be, for example, a vinylene group, a phenylene group, a methinimine
group, or combinations of those.

If the push-pull system is sufficiently large, we may split it into three parts, i.e., a left
(L), a central (C), and a right (R) part as shown in Fig.\ \ref{fig01}. Electrons of the central
part are assumed to be so far from the terminations that they do not feel the latter (or,
more precisely, the effects of the terminations are exponentially decaying in the central part). 

The dipole moment, $\vec\mu$, is useful in quantifying the response of the system to an 
external 
electric field,
\begin{eqnarray}
\mu_i(\omega)&=& \mu_i^{(0)}(\omega)  + \sum_j\sum_{\omega_1}\alpha_{ij}(\omega;\pm\omega_1) 
\cdot E_j(\omega_1)\nonumber\\  
&&+ \frac{1}{2}\sum_{jk}\sum_{\omega_1,\omega_2}\beta_{ijk}(\omega;\pm\omega_1,\pm\omega_2)\cdot 
E_j(\omega_1) E_k(\omega_2)\nonumber\\ 
&&+\frac{1}{6}\sum_{jkl}\sum_{\omega_1,\omega_2,\omega_3}\gamma_{ijkl}(\omega;\pm\omega_1,
\pm\omega_2,\pm\omega_3)
\cdot E_j(\omega_1) E_k(\omega_2) E_l(\omega_3) + \cdots.  
\label{eqn01}
\end{eqnarray}
Here, $E_m(\omega_s)$ is the $m$th component (i.e., $x$, $y$, or $z$) 
of the external field with the frequency $\omega_s$ and $\omega$ is the frequency of the response of 
the molecule to the field. The $\omega_n$ summations go over all 
the frequencies of the applied field. $\mu_i^{(0)}(\omega)$ is the dipole moment
in the absence of the field which vanishes for $\omega\ne 0$. Moreover, $\alpha_{ij}(\omega;\pm\omega_1)$ is 
the linear polarizability, and $\beta_{ijk}(\omega;\pm\omega_1,\pm\omega_2)$, 
$\gamma_{ijkl}(\omega;\pm\omega_1,\pm\omega_2,
\pm\omega_3)$, $\dots$ are the first, second, $\dots$ hyperpolarizability. Sum rules 
require that these quantities can be non-zero only if the frequency of the 
response, $\omega$, equals the sum of the frequencies (eventually multiplied by $-1$), i.e., 
for $\gamma_{ijkl}(\omega;\pm\omega_1,\pm\omega_2,\pm
\omega_3)$ we require $\omega=\pm\omega_1\pm\omega_2\pm\omega_3$.

In the present paper we focus on static external fields, in which case
$\omega_i=0$. Furthermore, we shall study a neutral system, although our arguments also are valid
for charged systems as long as the extra charge is localized to the terminations.
We let $\rho(\vec r)$ be the (field-dependent) total charge density (i.e., the sum of the nuclear and 
electronic charge densities), and choose the long axis to be $z$. Then the
component of the total dipole moment that is of interest here, namely $z$, is given by 
(omitting its argument, $\omega$)
\begin{equation}
\mu_z = \int \rho(\vec r)z d\vec r
=\int_L\rho(\vec r)z d\vec r+\int_C\rho(\vec r)z d\vec r+\int_R\rho(\vec r)z d\vec r,
\label{eqn02}
\end{equation}
where we have split the integral into contributions from the left, central,
and right regions of the chain. The central region consists of identical neutral units.
We can, therefore, write
\begin{equation}
\int_C\rho(\vec r)z d\vec r=K_C \mu_C,
\label{eqn03}
\end{equation}
where $K_C$ is the number of units in C and $\mu_C$ is the $z$ component of the dipole
moment of one of these units. In order to evaluate the other two contributions to the total
dipole moment in Eq.\ (\ref{eqn02}) we define a `typical' center for each term,
i.e., $\vec R_R$ and $\vec R_L$ (these could, e.g., be the center of mass
of the right and left parts, respectively), and let $Z_R$ and $Z_L$ be the $z$ 
components of these vectors. Since the chain is neutral we, then, obtain
\begin{equation}
\int_L\rho(\vec r)z d\vec r+\int_R\rho(\vec r)z d\vec r
=(Z_R-Z_L)\int_R\rho(\vec r)d\vec r + \int_L\rho(\vec r)(z-Z_L)d\vec r+
\int_R\rho(\vec r)(z-Z_R)d\vec r.
\label{eqn04}
\end{equation}
The first term on the right hand side describes the contribution to the dipole 
moment associated with electron transfer from one end to the other. This term grows linearly 
with chain length (due to $Z_R-Z_L$) as does the term in Eq.\ (\ref{eqn03}).
On the other hand, the last two terms in Eq.\ (\ref{eqn04}) describe local dipole
moments that arise from the electron distributions within the two terminal
regions and they are independent of the chain length.

This discussion suggests that donor/acceptor (=D/A) substitution at the ends of 
long chains may change the charge distribution in R and L so as to strongly
enhance the dipole moment and, consequently, produce a particularly large
change in the dipole moment when the system is exposed to an external electric field.
Therefore, very many studies have been devoted to push-pull systems as a function
of increasing length (see, e.g., 
[\onlinecite{mbz92,ty92,mgmpbbp94,sldzfss94,mtwc94,gpbfldbtncm94,hfmmzl95,m95,vtvg96,lmrfg96,bggpjm96,cjak97,zl99,stcm00,cpjgbsrk00,kcb00,bcta04,smrlls04}]).

Not only the electrons but also the
structure (phonons) will respond to a static electric field. We will  
demonstrate that, for sufficiently long chains, the electronic response per unit of a push-pull 
system (with structural relaxation taken into account) becomes
independent of the donor and acceptor groups, implying that the materials properties cannot be
improved upon substitution. Our mathematical arguments for this finding are presented in the 
next section, and in Sec.\ \ref{sec03} we illustrate and analyse the
results through calculations on a model system. The particular case of inversion symmetry
is discussed in Sec.\ \ref{sec04} where we also make a comparison with previous results.
Finally, a summary is provided in Sec.\ \ref{sec05}.

The arguments we present are related to those originally given by Vanderbilt 
and King-Smith for an extended system in the absence of an external field. They argued that
the permanent polarization (i.e.\ dipole moment per unit length)
is a bulk property.\cite{vks93} Very recently, Kudin {\it et al.}\cite{kcr07} 
proved that the permanent polarization is quantized for D/A substituted systems. 
Neither of these works considered the induced polarization or the 
structural relaxation due to an external field. Finally, in a recent paper we presented 
some of the arguments behind the present work but did not analyze the predictions 
as we do here using a model system.\cite{gut}

\section{Changes in the Charge Distribution upon Substitution}
\label{sec02}

By replacing some (groups of) atoms with others at the chain ends,
the electronic orbitals with components near the ends will change. 
Since the set of electronic orbitals is orthonormal, all other orbitals 
will change as well. Accordingly, the charge distribution may change everywhere 
due to the substitutions.

When an electrostatic field is applied as well, each orbital will respond to 
the field. Since the orbitals will have changed due to the substitution, so will
their responses to the field. Furthermore, the structural responses due to the field will also depend
on the substitution at the ends. Therefore, the dipole moment can depend upon both the 
substitution and the field. From these arguments there is no reason to believe that
$\mu^{(0)}/N$, $\alpha/N$, $\beta/N$, $\gamma/N$, $\dots$ (with $N$ being the number of repeated units)
will be independent of the substitution. However, we shall argue here that the charge
\begin{equation}
q =  \int_R\rho(\vec r)d\vec r
\label{eqn05}
\end{equation}
in Eq.\ (\ref{eqn04}) can change, at most, by an integral number of elementary 
units for different D/A substitutions at fixed external static
field. Our proof is a generalization
of arguments due to Vanderbilt and King-Smith\cite{vks93} (see also [\onlinecite{kcr07}]),
and was previously proposed by the present authors.\cite{gut} It will be verified 
here by calculations on a model system and given a thorough analysis on that 
basis.

For a given system (with specified geometry), and value of the external field, we transform the set of occupied 
orbitals into a set of orthonormal, localized functions. Those functions 
ascribed to C will be similar to the Wannier functions of the infinite 
periodic system. The localized orbitals will be centered
in one region, but may have tails sticking into another region. 
We assume that the terminal regions are large enough so that any functions centered 
therein, which differ from those of C, are exponentially vanishing in C. 
On the other hand, those functions ascribed to C, but centered on 
units next to L or R,
will likely have tails extending into those regions. 

The density matrix can then
be written in block-diagonal form with three blocks, one for each of the three regions.
Since the density matrix is idempotent, each block will be so, too, and 
there will be an integral number of electrons 
associated with each of the three sets of functions. That is to say, the number of 
electrons associated with the functions centered in the two end regions is integral. 
Accordingly, any non-integral part of $q$ is associated with the tails of the 
functions in C that extend into R, which, per construction, is independent of the
terminations, i.e., also of D/A substitution.

We conclude that, for different terminations, $q$ can change only by an integer. This is valid
for long chains and all fields. Therefore, the electronic response per unit of the chains to
the field, with or without nuclear response, is independent of termination. The only possible
change for different terminations is that $q$ may jump by an integer for different field 
strengths. In fact, our numerical studies on a H\"uckel-type model will confirm this 
prediction. Of course, in ab initio calculations, there may also be a jump due to changing the
basis set or the method (e.g.\ Hartree-Fock vs. Kohn-Sham DFT).

\section{Illustrating and Analyzing the Result}
\label{sec03}

In order to explore in detail the predictions from above we studied a H\"uckel like model for 
long, finite (AB)$_{2K+1}$ chains. In our model, we use a basis set of orthonormal 
atomic orbitals (AOs) with one AO per atom. The system has one electron per atom, and
the nuclei are given charges of $+1$ whereas the electronic charge is set equal to $-1$. (All
quantities are expressed in atomic units in this paper.) 
Given that $\chi_n$ is the AO of the $n$th atom ($n=1,2,\dots,4K+2$) and $\hat h$ is
the Kohn-Sham or Fock single-electron hamiltonian we assume that only $\langle \chi_j\vert\hat h\vert \chi_j\rangle$,
$\langle \chi_j\vert\hat h\vert \chi_{j\pm1}\rangle$, and 
$\langle \chi_j\vert\hat h\vert \chi_{j\pm2}\rangle$ are non-vanishing with values
\begin{eqnarray}
\langle \chi_{2p+1}\vert\hat h\vert \chi_{2p+1}\rangle&=&\epsilon_0\nonumber\\
\langle \chi_{2p}\vert\hat h\vert \chi_{2p}\rangle&=&-\epsilon_0\nonumber\\
\langle \chi_j\vert\hat h\vert \chi_{j+1}\rangle&=& -[t_1-\alpha_1(z_{j+1}-z_j)]\nonumber\\
\langle \chi_j\vert\hat h\vert \chi_{j+2}\rangle&=& -[t_2-\alpha_2(z_{j+2}-z_j)]. 
\label{eqn06}
\end{eqnarray}
Here $z_j$ is the position of the $j$th atom. Different donor and acceptor groups are 
modeled by modifying the on-site energies of the terminating atoms and/or the terminating 
hopping integrals, 
\begin{eqnarray}
\langle \chi_1\vert\hat h\vert \chi_1\rangle&=&\epsilon_0+\epsilon_L\nonumber\\
\langle \chi_{4K+2}\vert\hat h\vert \chi_{4K+2}\rangle&=&-\epsilon_0+\epsilon_R\nonumber\\
\langle \chi_1\vert\hat h\vert \chi_2\rangle&=& -[t_1-\alpha_1(z_2-z_1)]+t_L\nonumber\\
\langle \chi_{4K+1}\vert\hat h\vert \chi_{4K+2}\rangle&=& -[t_1-\alpha_1(z_{4K+2}-z_{4K+1})]+t_R. 
\label{eqn07}
\end{eqnarray}

Finally, we assume that 
\begin{equation}
\langle\chi_j\vert z\vert\chi_k\rangle=\delta_{j,k}z_j.
\label{eqn07a}
\end{equation}

In order to analyse the results we, first, define a reference structure for which the position of the 
$n$th atom is 
\begin{equation}
z_n^{(0)}=\frac{a}{2}\left(n-2K-\frac{3}{2}\right)-(-1)^nu_0.
\label{eqn08}
\end{equation}
Here $a$ is the length of the unit cell for an infinite, periodic system with
the same electronic interactions and no external field. 
Subsequently, we define for each atom
\begin{eqnarray}
u_n&=&z_n-\frac{a}{2}\left(n-2K-\frac{3}{2}\right)\nonumber\\
\Delta z_n &=& z_n-z_n^{(0)}.
\label{eqn09}
\end{eqnarray}

The total energy is written as the sum over occupied orbital energies (multiplied by 2 due to spin degeneracy) 
augmented by a harmonic term in the nearest- and the next-nearest-neighbour bond lengths,
\begin{equation}
E_{\rm tot}=2\sum_{i=1}^{\rm occ}\epsilon_i + 
\frac{k_1}{2}\sum_{p=1}^{4K+1}(z_{p+1}-z_p)^2+
\frac{k_2}{2}\sum_{p=1}^{4K}(z_{p+2}-z_p)^2-E_{\rm DC}\sum_{p=1}^{4K+2} z_p.
\label{eqn09a}
\end{equation}
$E_{\rm DC}$ is the strength of the electrostatic field.
For the infinite, periodic chain without an external field, the lowest total energy
corresponds to a certain lattice constant $a$ and  
\begin{equation}
u_n=(-1)^{n+1} u_0. 
\label{eqn09b}
\end{equation}
The force constants $k_1$ and $k_2$ are determined so that $a$ and $u_0$ take certain 
chosen values.

With 
\begin{equation}
\psi_i = \sum_{n=1}^{4K+2} C_{ni} \chi_n
\label{eqn10}
\end{equation}
being the $i$th orbital (ordered according to increasing orbital energy) we calculate 
the Mulliken charge on the $n$th atom for field $E_{\rm DC}$ as 
\begin{equation}
q_n(E_{\rm DC}) = 1 -2\sum_{i=1}^{2K+1} \vert C_{ni}\vert^2
\label{eqn11}
\end{equation}
which leads to the dipole moment
\begin{equation}
\mu_z = \sum_{n=1}^{4K+2} z_n q_n(E_{\rm DC}).
\label{eqn12}
\end{equation}
The charge transfer is given through
\begin{equation}
q = \sum_{n=2K+2}^{4K+2} q_n(E_{\rm DC}).
\label{eqn13}
\end{equation}
We also define
\begin{eqnarray}
\Delta_1 q_n(E_{\rm DC}) &=& q_n(E_{\rm DC})-\tilde q_n(0)\nonumber\\
\Delta_2 q_n(E_{\rm DC}) &=& q_n(E_{\rm DC})-q_n(0).
\label{eqn14}
\end{eqnarray}
where $\tilde q_n(0)$ is the charge for the infinite, periodic chain in the absence of 
the field. $\Delta_2 q_n(E_{\rm DC})$ quantifies the effects on the charge distribution of
the push-pull chain due to
including the field, whereas $\Delta_1 q_n(E_{\rm DC})$ includes effects both from the field
and from the terminations. Note that $\Delta_1 q_n(E_{\rm DC})-\Delta_2 q_n(E_{\rm DC})$ gives the
field-independent effect of the terminations. Finally, it turns out to be useful to define
the center and width of the $i$th orbital according to
\begin{eqnarray}
\zeta_i &=& \sum_{n=1}^{4K+2}z_n \vert C_{ni}\vert^2\nonumber\\
\Delta\zeta_i &=& \left[\sum_{n=1}^{4K+2}(z_n-\zeta_i)^2 \vert C_{ni}\vert^2\right]^{1/2},
\label{eqn15}
\end{eqnarray}
which is consistent with Eq.\ (\ref{eqn07a}). 

\begin{figure}
\begin{picture}(15,17)
\put(0,0){\psfig{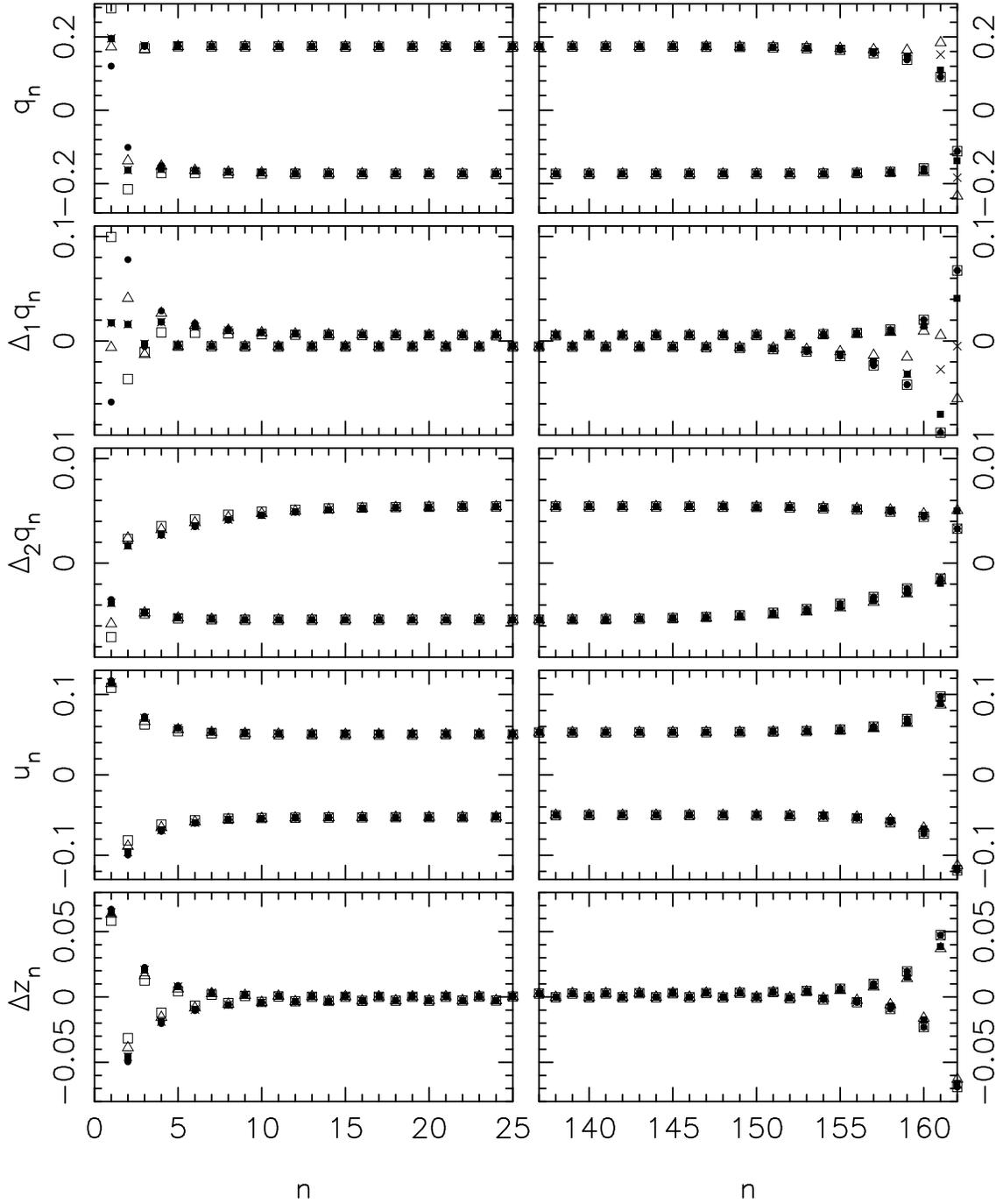}}
\end{picture}
\caption{Results from the model calculations for chains with $K=40$, i.e., 162 atoms 
and an external field of -0.015 a.u.
The first three rows show quantities related to the
charge distribution, i.e., those of Eqs.\ (\ref{eqn11}) and (\ref{eqn14}), whereas
the two lowest rows show quantities related to the structure, i.e., those of 
Eq.\ (\ref{eqn09}). In all cases, $\epsilon_0=0.5$, $t_1=2.5$, $\alpha_1=4.5$,
$t_2=0.5$, and $\alpha_2=0.2$. Moreover, $k_1$ and $k_2$ were chosen so that 
the optimized structure for vanishing field and an infinite periodic chain 
has $a=2.1$ and $u_0=0.05$. Open circles, closed circles, crosses, open squares,
closed squares, and open triangles show results for which 
$(\epsilon_L,\epsilon_R,t_L,t_R)$ has been set equal to $(0,0,0,0)$, 
$(0,0.5,0,0)$, $(0.5,0,0,0)$, $(0,0,0,0.8)$, $(0,0,0.8,0)$, and $(0.5,0.5,0.8,0.8)$,
respectively.}
\label{fig02}
\end{figure}

\begin{figure}
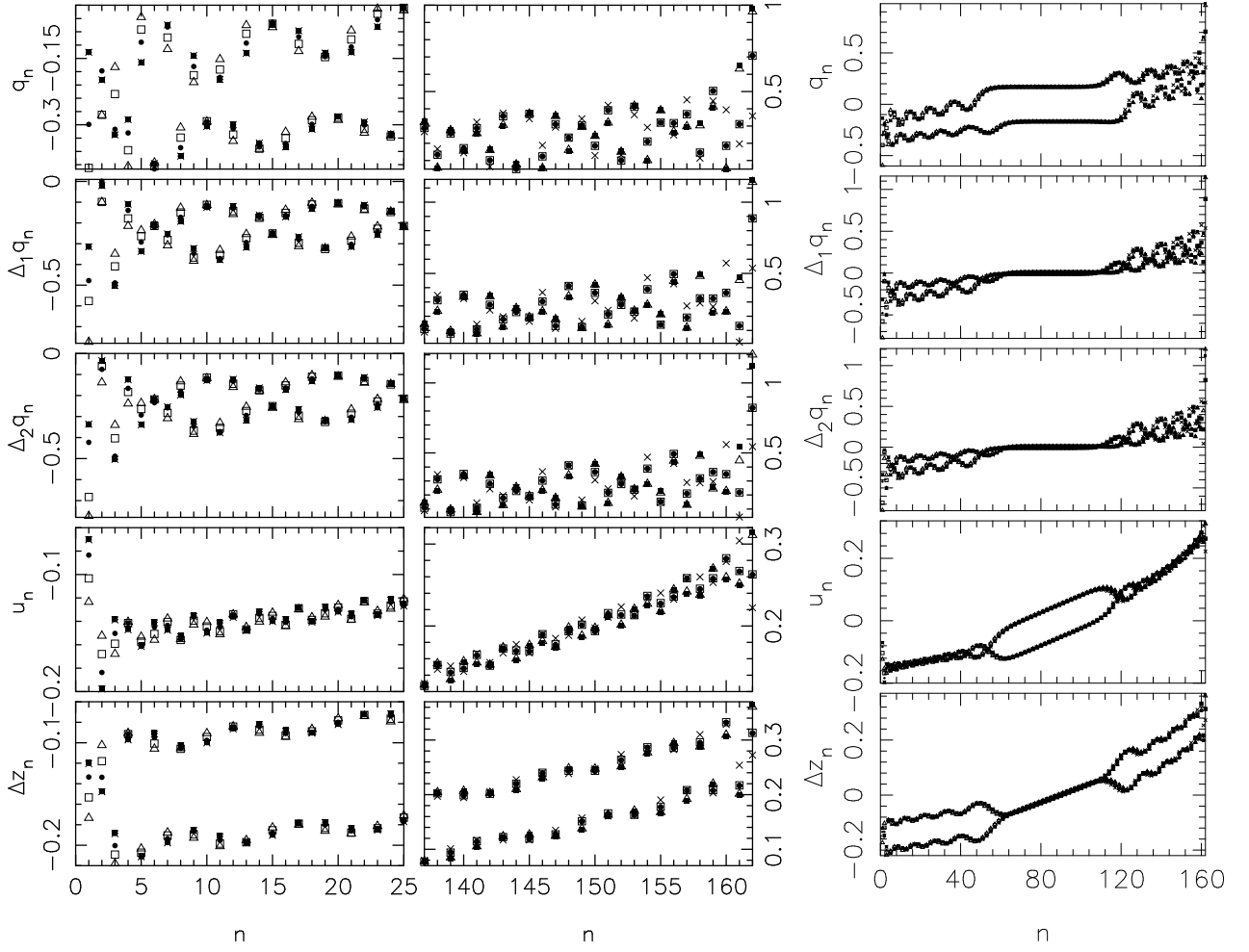

\begin{picture}(15,13)
\put(-0.5,0){\psfig{file=zzz2.ps,height=13cm}}
\put(10.5,0){\psfig{file=zzz3.ps,height=13cm}}
\end{picture}
\caption{Results from the model calculations for chains with $K=40$, i.e., 162 atoms 
and an external field of -0.03 a.u.
The left and middle columns show results for the two ends whereas
the right column shows the results for the whole chain. The presentation is as
in Fig.\ \ref{fig02}.}
\label{fig02a}
\end{figure}

\begin{figure}
\begin{picture}(15,15)
\put(00,0){\psfig{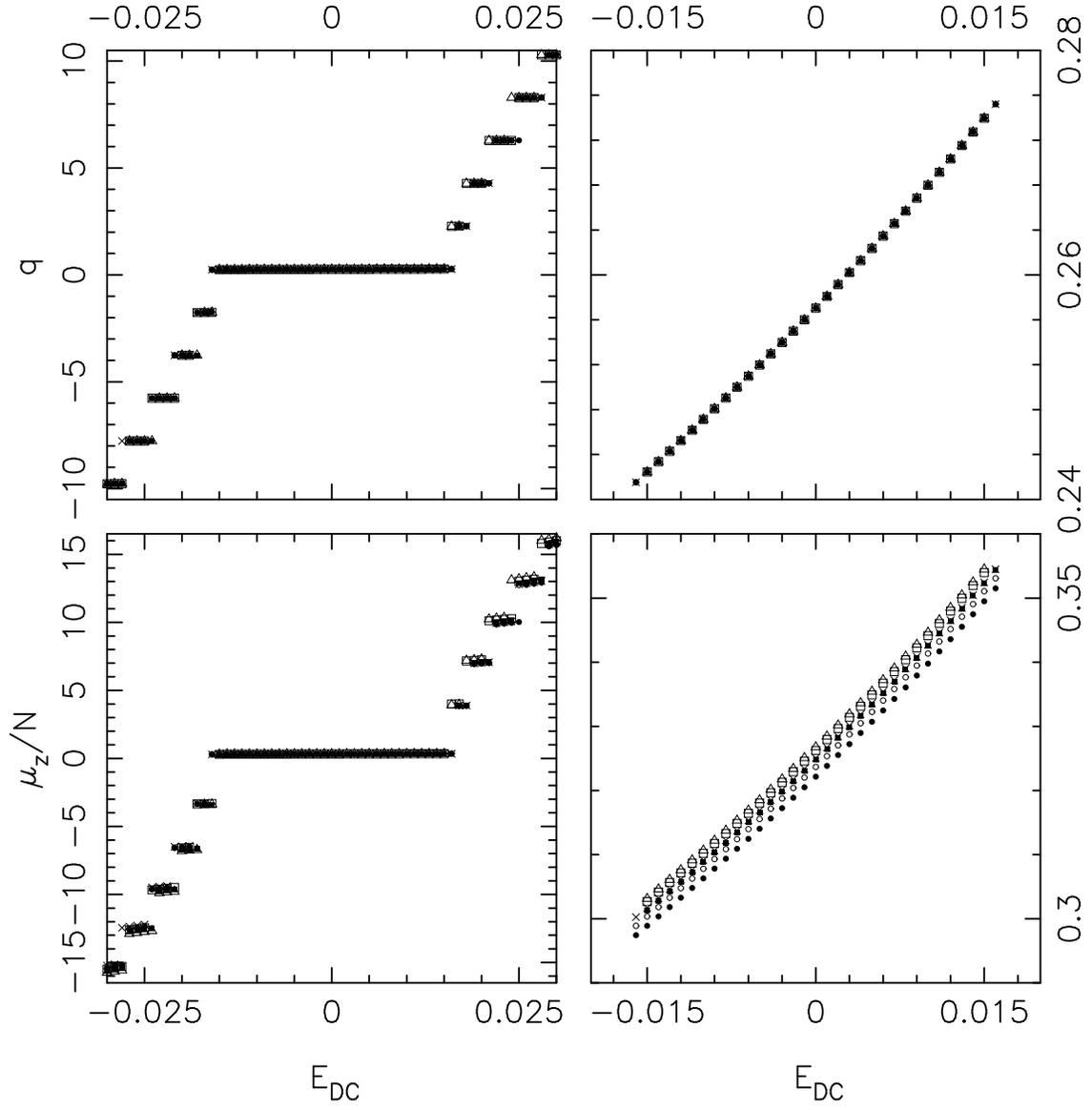}}
\end{picture}
\caption{The charge transfer, $q$, and the dipole moment per unit, $\mu/N$ with $N=2K+1$,
for the same systems as in Fig.\ \ref{fig02}. The right panels show 
a magnification of the left ones.}
\label{fig04}
\end{figure}

\begin{figure}
\begin{picture}(15,20)
\put(00,0){\psfig{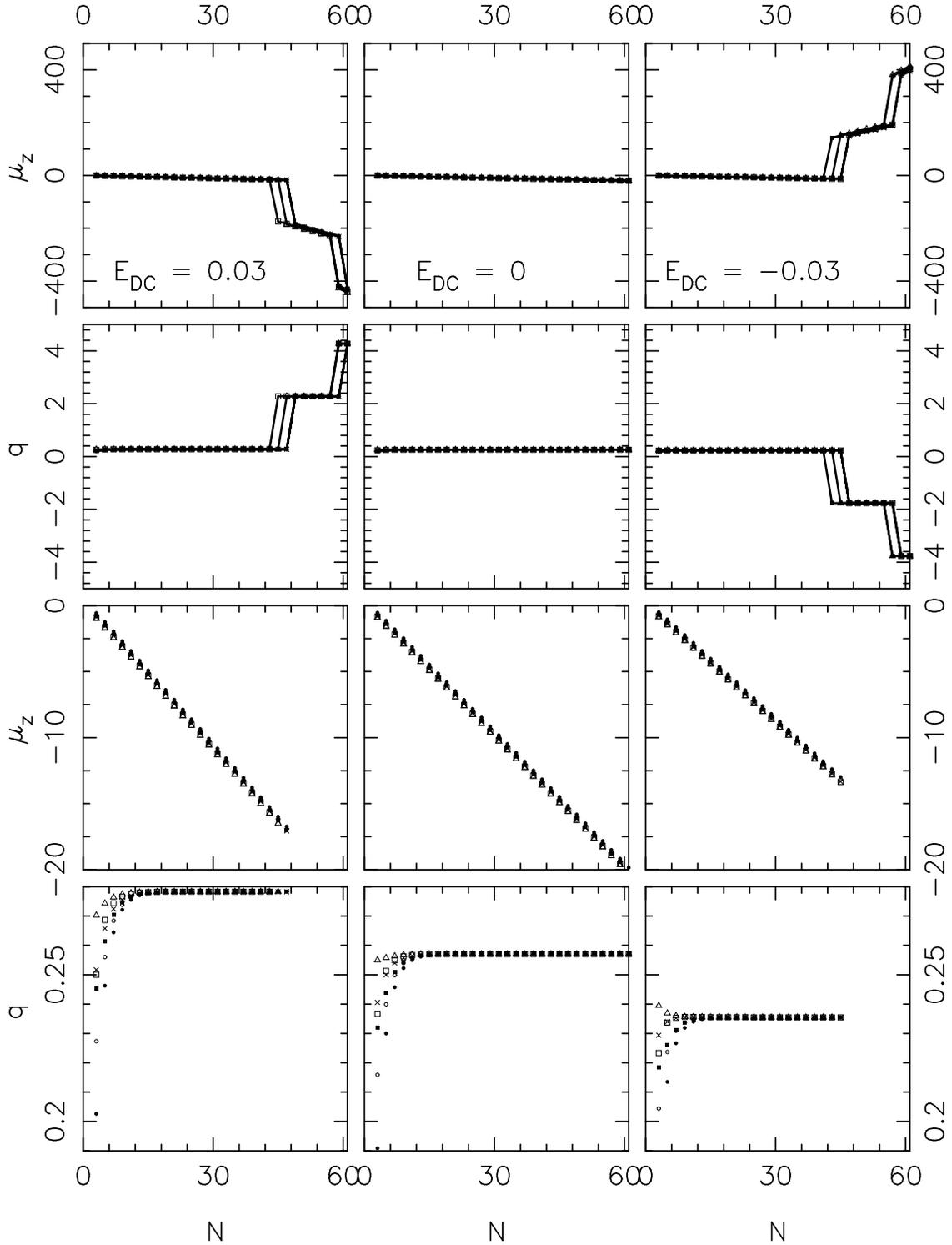}}
\end{picture}
\caption{The charge transfer, $q$, and the dipole moment $\mu$ with $N=2K+1$,
for the same systems as in Fig.\ \ref{fig02} as functions of $N$. The two lower rows show 
a magnification of the two upper ones. The field strength has been set equal to $0.03$, 0,
and $-0.03$ for the left, middle, and right columns, respectively.}
\label{fig05}
\end{figure}

\begin{figure}
\begin{picture}(15,15)
\put(00,0){\psfig{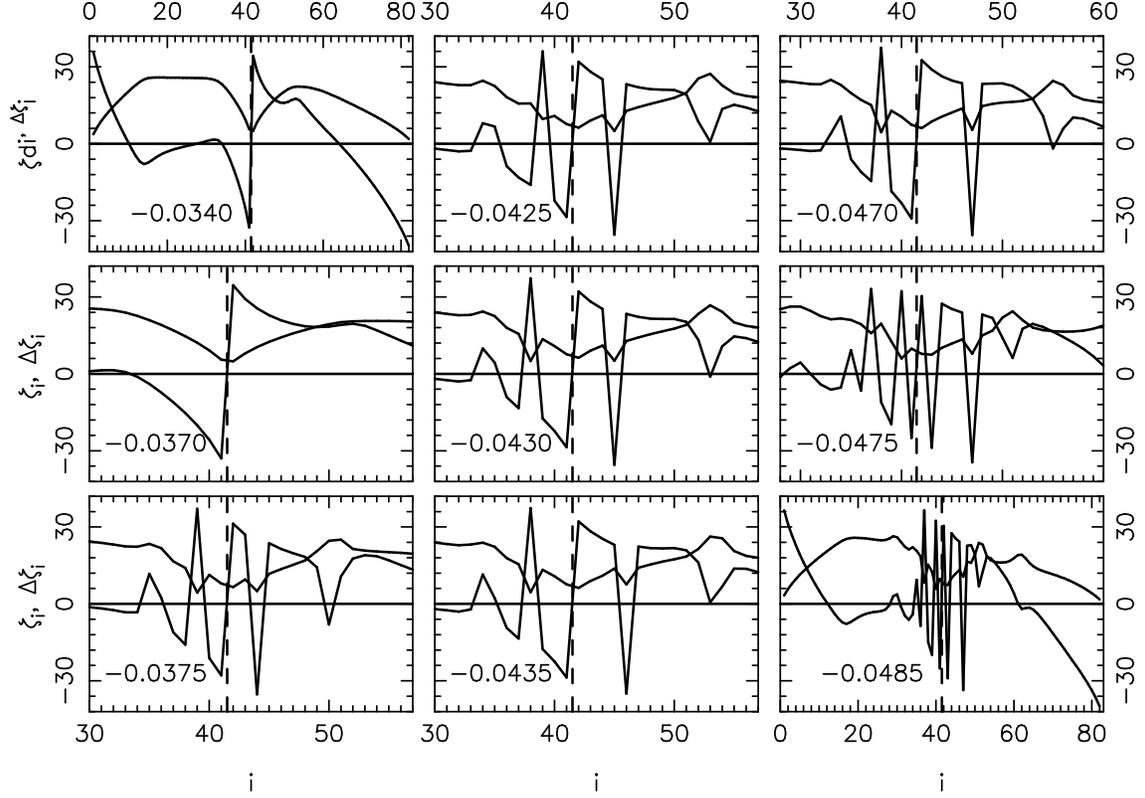}}
\end{picture}
\caption{$\zeta_i$ and $\Delta\zeta_i$ as a function of orbital index for a given push-pull chain with
different fields. The 
vertical dashed lines separate occupied and unoccupied orbitals; $K=20$; and the
parameter values are the same as in Fig.\ \ref{fig02} except that only the first
termination is considered. The field strength is given in each panel. Since
$\Delta\zeta_i\ge0$ the two curves are readily distinguished. Moreover, the panels for
$E_{\rm DC}=-0.0340$ and $-0.0485$ have the same orbital index scales,
as is the case for the panels for $E_{\rm DC}=-0.0370$, $-0.0375$, $-0.0425$, $-0.0430$, and $-0.0435$, and
for the panels for $E_{\rm DC}=-0.0470$ and $-0.0475$.}
\label{fig06}
\end{figure}

We performed calculations for six different 
terminations specified by $(\epsilon_L,\epsilon_R,t_L,t_R)$.
The results are summarized in Figs.\ \ref{fig02}, \ref{fig02a}, \ref{fig04}, \ref{fig05}, and \ref{fig06}.
Since our model is that of a finite chain with two different types of atoms, A and B, the 
Mulliken charges in the central region take two values. This is clearly recognized in the
presentation of $q_n$ in Fig.\ \ref{fig02} for $E_{\rm DC}=-0.015$. 
In Fig.\ \ref{fig02} it is also seen that near the ends,
the Mulliken charges differ from the values of the inner part and, moreover, these charges
depend sensitively on the terminations. For the field strength $E_{\rm DC}=-0.015$
these findings are only marginally modified compared to those of a vanishing field 
(not shown). From $\Delta_1 q_n$ we see that the combination of electrostatic
field and termination leads to an internal polarization of each unit in C. 
Actually, $\Delta_1 q_n$ shows a reduced internal polarization compared to $\Delta_2 q_n$. Thus,
terminating the chain reduces the effect of the field in that regard. Whereas $\Delta_2 q_n$
contains information about the field-induced charge
redistributions, $\Delta_1 q_n$ contains additional information about the (field-dependent)
effects of the terminations. For $E_{\rm DC}=-0.015$  
the field-induced charge redistributions are smaller near the terminations than 
in the central parts. 

For the larger field, $E_{\rm DC}=-0.03$, in Fig.\ \ref{fig02a} 
the identification of the central region becomes much more difficult and, as
we shall see below, electrons are transferred from one end to the other. Moreover, 
in this case the field perturbs the system so strongly that the effects of the field are stronger
than those of the terminations. This can be seen from the fact that $\Delta_1 q_n$ and 
$\Delta_2 q_n$ are very similar. 

The structure also depends upon the termination. For the intermediate field
of Fig.\ \ref{fig02} (and for zero field as well) the atomic coordinate $u_n$ is nearly 
constant in C but varies considerably near the ends where its value depends on the 
termination, as was the case for the atomic charges. 
For the higher field in Fig.\ \ref{fig02a} it appears as if no 
central region can be identified from this parameter. However, the fact that $\Delta z_n$ is 
essentially linear for the innermost atoms implies that there is a well-defined,
repeating structure in C with a lattice constant differing from that of the field-free 
case.

Fig.\ \ref{fig04} shows that the charge transfer, $q$, is independent of termination (though
not independent of the field), with the
exception of jumps by (even) integers. (The integers are even because we have not allowed 
for spin polarization.) However, the charge distribution inside  
R or L does depend on the terminations and, as a consequence, the 
dipole moment does as well. On the other hand, the variation of 
$\mu_z/N$ as a function of $E_{\rm DC}$ for different terminations follows parallel curves, 
implying that the (hyper)polarizabilities are independent of the terminations. In fact, 
a least squares fit yields the values (including maximum deviations): $\mu_0/N = 0.3245\pm
0.0023$, $\alpha/N = 1.677\pm 0.013$, $\beta/N = 18.17 \pm 0.20$, and 
$\gamma/N = 606.9 \pm9.1$  for all six terminations.

As a function of field $\mu_z$ is discontinuous and the power series expansion
is valid only up to the field where the discontinuity occurs. Once such a 
discontinuity has been passed, the dipole moment depends more strongly on the field. This 
means that the only way of increasing the responses of long push-pull systems to DC 
fields is to design chains for which the integral electron transfers occur at low fields.

At a given field the size of the chain for which jumps in the charge $q$ (i.e.\ Zener
tunneling) take place depends on the terminations (cf.\ Fig.\ \ref{fig05}). In the 
shortest chains, for which Zener tunneling does not occur, $\mu_z$ follows parallel 
curves as a function of chain length, $N=2K+1$, for different 
terminations. This means that the 
dipole moment and (hyper)polarizabilities per unit become 
independent of termination. However, as seen in Fig.\ \ref{fig05}, the slope of these curves
increase after Zener tunneling has taken place, implying that the dipole moment 
increases. Assuming that the field-dependence of the dipole moment likewise increases, this suggests
that the polarizability and/or hyperpolarizabilities per unit may increase for D/A
substituted systems after an integral number of electrons has been transferred from one end to 
the other. 

In Fig.\ \ref{fig06} we show an example of what happens to the molecular orbitals
when the jumps take place. Calculations were performed for field strengths between
$-0.0340$ and $-0.0485$ in steps of $-0.0005$, but in the figure we only show the results
for fields where Zener tunneling occurs. In all cases, the curves vary smoothly as a function
of field strength. 
At the lowest two fields, the occupied orbitals closest to the Fermi level have a center
in the left part ($\zeta_i<0$), whereas the unoccupied orbitals closest to the Fermi level 
are centered in the right part. At the field $E_{\rm DC} \simeq -0.0375$, two 
electrons (one per spin direction) are 
transferred from one side to the other, which again happens at a larger field 
($E_{\rm DC} \simeq -0.0475$).
In the first case, we observe the occurrence of two new, very localized, orbitals 
close to
(but not at) the Fermi level. The energetically lower (i.e occupied) one is 
localized towards to the chain end on the right side while the other (unoccupied)
is localized towards the chain end on the left side. Accompanying this interchange is 
a similar interchange of two rather delocalized 
orbitals, both of which are further away from the Fermi level and centered closer to the
middle of the chain.
Again, at the second electron transfer a pair of new, rather localized, orbitals
near (even closer to) the Fermi level show up towards the chain ends,
and also this transfer is accompanied by some reorganization of the other orbitals. 
Finally, Fig.\ \ref{fig06} also shows an example of a reorganization of the orbitals, i.e., 
for a field around $E_{\rm DC} = -0.0430$. Here, one localized, occupied orbital interchanges
order with an adjacent (in energy) more delocalized orbital, but otherwise no further 
significant changes are observed.

\section{Inversion symmetry and comparison with previous results}
\label{sec04}

Before proceeding to compare with previous results we develop an interesting consequence
of our findings with regard to inversion symmetry. The same arguments can be applied for 
a system containing a mirror plane perpendicular to the 
chain axis, but here we shall for the sake of simplicity restrict ourselves to the 
case of inversion symmetry. Suppose the
long oligomer of interest contains a central region made up of units with inversion
symmetry. Even if the central part does not have inversion symmetry, it may be 
possible to create such with the addition of appropriate terminating groups. This 
is, for example, the case for oligomers of thienyleneethynylenes and thienylenevinylenes that were
studied by Geisler {\it et al.}\cite{gpbfldbtncm94} Many of the systems of interest fall 
into one of these two categories. Since, according to our findings, D/A substitution 
cannot affect the (hyper)polarizabilities per unit, the latter must vanish even if the symmetry is not 
preserved. For instance, modifying the terminations of the systems of Geisler {\it et al.} so 
that inversion symmetry no longer exists cannot result in a non-vanishing $\beta/N$ if the chains
are sufficiently long.

A large fraction of previous observations are for systems of the type 
described in the preceding paragraph. Some of these cases are discussed below along
with others pertinent to our findings herein. We now briefly consider, in particular,
the works mentioned in the Introduction.

In their combined experimental and theoretical study
on some push-pull oligoenes, Meyers {\it et al.}\cite{mbz92} observed a `negligible charge 
transfer all the way from the donor to the acceptor', which implies that $q$ is independent 
of the termination. On the other hand, in their theoretical study Tsunekawa and 
Yamaguchi\cite{ty92} examined shorter, nitrogen-containing push-pull oligomers. They noted that
these systems are interesting from the perspective of maximizing $\beta$, but our results 
establish that, for such to be true, the systems must be  short enough so that our approach 
is inapplicable. This serves to highlight the point that apparent, but not real, 
discrepancies can occur due to shortness of the chain length.

Marder {\it et al.}\cite{mgmpbbp94} presented an approach for unifying the description of linear 
and nonlinear polarization in organic polymethine dyes. It has since been shown that their 
analysis is invalid if phonons are taken into account.\cite{kcb00} Here, however, we 
emphasize that the conclusions they draw regarding $\beta$ can, again, hold only for 
systems that are too short for our treatment to apply.

Clearly, the chain length required for validity of the treatment given here is an important
issue. In Fig.\ \ref{fig05} the dipole moment is converged for chains with some 20 units. 
However, this may be an artifact of our simple H\"uckel model. In an experimental 
study\cite{sldzfss94} and in several computational studies,\cite{mtwc94,vtvg96,lmrfg96,stcm00,ktrh95}
the second hyperpolarizability per unit was found to converge considerably slower which, 
in fact, agrees with our own earlier findings.\cite{gut} Thus, when focusing on 
higher-order non-linear responses quite large chains may be required for 
the results of the present work to be relevant. In
shorter push-pull systems (for instance those considered
by Geisler {\it et al.}\cite{gpbfldbtncm94,bggpjm96} or by 
Morley {\it et al.}\cite{hfmmzl95,m95}) D/A substitution can 
have an influence on the response.

As shown numerically by Champagne {\it et al.},\cite{cjak97} $\beta/N$ also
converges relatively slowly as a function of size. They considered D/A 
substituted oligomers of polymetheimine [also called polycarbonitrile, (CHN)$_x$]. This 
system has a zigzag backbone of alternating C and N atoms with alternating bond lengths. 
Without the bond length alternation it would, at least hypothetically, be possible to choose 
donor and acceptor groups so that the overall system is centrosymmetric. 
Even if chemical arguments imply that this structure is unrealistic, a non-zero value of
$\beta/N$ for long chains should be ascribed, strictly speaking, to the bond length 
alternation.

Polyphenylenes and polypyridines have been studied by 
Zhang and Lu.\cite{zl99} They focused on $\alpha$ and $\gamma$ as a function of 
the length of a closed
ring for each system and applied a finite-field approach in their calculations. 
Unfortunately, as we have shown earlier (see, e.g., [\onlinecite{gut}]), this approach will never
converge to the results for the infinite, periodic chain. Nevertheless, although $\beta/N$
will vanish for the polyphenylenes, we predict that a non-zero value will occur
for both short and long oligomers of the polypyridines.

For the D/A substituted polyenes studied by Champagne {\it et al.}\cite{cpjgbsrk00} our
analysis confirms their findings, i.e., that 
$\beta/N$ will vanish for sufficiently large chains. Their numerical results indicate that 
$\beta/N$ goes through a maximum and that convergence to the infinite chain result for 
larger $N$ is slow. 

Even the polarizability, $\alpha/N$, and the permanent dipole moment, $\mu_z^{(0)}/N$, 
may converge more slowly, as a function of chain length, than
predicted by our simple model. This is, for example, the case for the systems
investigated by Smith {\it et al.}\cite{smrlls04} and by Kudin {\it et al.}\cite{kcr07}

In a recent study, Botek {\it et al.}\cite{bcta04} compared finite oligomers of [$N$]helicenes 
and [$N$]phenylenes that possess a helical structure for $N$ larger than roughly 6. By making 
explicit use of the helical symmetry of the central region we predict that, when those 
systems are sufficiently long, D/A substitution will not be able to modify the electronic 
responses to static fields. The fact that 
Botek {\it et al.} find changes upon D/A substitution implies that the
chains of their study are not converged to the long chain limit.

\section{Summary}
\label{sec05}

As long as the applied field is not so strong that an integral number of electrons is 
transferred from one end to the other, the answer to the question of 
the title is clearly: there can be no change. This comes from our mathematical 
analysis in Sec.\ \ref{sec02}, which generalizes treatments presented previously by Vanderbilt
and King-Smith\cite{vks93} and by Kudin {\it et al.},\cite{kcr07} who considered only 
electronic polarization in the absence of an external electrostatic field. It is also in 
agreement with our own earlier prediction.\cite{gut}

Calculations on a model system confirm the basic result and shed light on the nature of
the end-to-end charge transfer. Although the end charges, permanent dipole moment, and 
structure depend sensitively on the terminations neither the amount of charge transferred nor the 
(hyper)polarizabilities per unit do so. The field and/or chain length at which the charge jumps take 
place also depend on the terminations. Each jump is associated with an interchange of 
occupied and unoccupied molecular orbitals that are well-localized in the chain end 
region. These orbitals are close to but not at the Fermi level. There is also an
accompanying orbital reorganization. 

One consequence of our finding is that long unsubstituted chains which have inversion or mirror
symmetry, or can be made symmetric by substitution, must have a vanishing first 
hyperpolarizability per unit. Experimental and theoretical determinations are consistent with this 
fact, although apparent contradictions can occur for short chains.

\begin{acknowledgments}

This work was supported by the German Research Council (DFG) through project
Sp439/20 within the SPP 1145. Moreover, one of the authors (MS) is very grateful to the 
International Center for Materials Research, University of California, 
Santa Barbara, for generous hospitality.

\end{acknowledgments}

\newpage

\unitlength1cm

\unitlength1cm

\unitlength1cm

\unitlength1cm

\unitlength1cm

\unitlength1cm

\end{document}